# Infrared-transmittance tunable metal-insulator conversion device with thin-film-transistor-type structure on a glass substrate

Takayoshi Katase[1,2,a),b)], Kenji Endo[3], and Hiromichi Ohta[1,c)]

[1]*Research Institute for Electronic Science, Hokkaido University, N20W10, Kita, Sapporo 001−0020, Japan*

[2]*PRESTO, Japan Science and Technology Agency, 5 Sanbancho, Chiyoda, Tokyo, 102−0075, Japan*

[3]*Graduate School of Information Science and Technology, Hokkaido University, N14W19, Kita, Sapporo 060−0814, Japan*

[a)] Present address: Laboratory for Materials and Structures, Institute of Innovative Research, Tokyo Institute of Technology, 4259 Nagatsuta, Midori, Yokohama, 226−8503, Japan

[b)] katase@es.hokudai.ac.jp

[c)] hiromichi.ohta@es.hokudai.ac.jp

Infrared (IR) transmittance tunable metal-insulator conversion was demonstrated on glass substrate by using thermochromic vanadium dioxide ($VO_2$) as the active layer in three-terminal thin-film-transistor-type device with water-infiltrated glass as the gate insulator. Alternative positive/negative gate-voltage applications induce the reversible protonation/deprotonation of $VO_2$ channel, and two-orders of magnitude modulation of sheet-resistance and 49% modulation of IR-transmittance were simultaneously demonstrated at room temperature by the metal-insulator phase conversion of $VO_2$ in a non-volatile manner. The present device is operable by the room-temperature protonation in all-solid-state structure, and thus it will provide a new gateway to future energy-saving technology as advanced smart window.

Thermochromic vanadium dioxide ($VO_2$) exhibits a reversible metal-insulator (MI) transition at a critical temperature ($T_{MI}$) of 68 °C due to the structural and electronic structure changes.[1-3] Above $T_{MI}$, $VO_2$ has a rutile-type tetragonal structure that is a metal and reflective to infrared (IR) due to the free carriers. In the low-temperature phase below $T_{MI}$, the V ions form a dimer, resulting in a monoclinic structure.[4] This structural transformation accompanies a dramatic change in the 3d-band configuration with appearance of charge gap ~0.6 eV, where the $VO_2$ changes to be an electrical insulator and transparent to IR.[5] Thus $VO_2$ has the potential to demonstrate IR-transmittance tunable MI conversion device on a glass substrate (Fig. 1), which is expected to have a significant contribution on energy saving technology as an advanced smart window: e.g. the device can selectively regulate thermal radiation from sunlight, and function as ON/OFF power switch to control the in-house temperature, which thus greatly reduces energy consumption including light expenses and cooling/heating loads.

For the application to such device, the control of $T_{MI}$ in the boundary of ambient temperature is inevitable to modulate the MI transition charecteristics of $VO_2$. Electron doping by the substitution of aliovalent metal ions, such as tungsten, at V site in $VO_2$ can reduce the $T_{MI}$ down to below room temperature (RT),[6] but it is an irreversible process. The proton doping (protonation) is one of the effective ways to reversibly modulate the $T_{MI}$; the proton can exist at interstial site in $VO_2$ lattice and acts as shallow donor that donates an electron to V ion,[7] similar to the electrochromic effect in $H_xWO_3$.[8] Compared to state-of-the-art modulation techniques, such as electrostatic-charge doping[9,10] and epitaxial strains in thin films,[11] the protonation of $VO_2$ ($H_xVO_2$) via a chemical route is the most ideal to switch the MI transition due to the intrinsic non-volatile operations.

However, there have been technological difficulties in the protonation of $VO_2$, which required a high-temperature heating with hydrogen source.[12-14] The most appropriate route for the RT protonation is electrochemical reaction in an electrolyte, but the uptake amount of proton ($H^+$) into $VO_2$ was too small to modulate the MI transition, where the proton insertion proceeded only to a value of $x = 0.06$ in $H_xVO_2$,[12] and the liquid-electrolyte likely causes a leakage problem that limits the application in practical use. Then we have recently demonstrated a new approach of water-electrolysis-induced protonation for $VO_2$ epitaxial film grown on sapphire substrate by using a three-terminal TFT-type structure with water-infiltrated nanoporous glass as a gate insulator;[15,16] this device is kind of a pseudo-solid-state electrochemical cell with a nano-gap parallel plate structure composed of $VO_2$ channel and metal gate electrode, where a gate bias application induces water electrolysis in the gate insulator and the produced $H^+$ / hydroxyl ($OH^-$) ion can be used to protonate / deprotonate the $VO_2$ channel, resulting in the reversible MI phase modulation at RT.[16] This RT-protonation approach can be realized in an all-solid-state TFT-type structure, and thus the advantageous feature should lead to a practical IR-transmittance tunable MI conversion device on a glass substrate.

Herein, we demonstrate an IR-transmittance tunable MI conversion device by extending the three-terminal TFT-type structure with water-leakage-free gate insulator to a large-area $VO_2$ film prepared on a glass substrate. **Figure 2(a)** schematically illustrates the device structure, which has a typical three-terminal TFT geometry composed of an active $VO_2$ layer, a gate insulator, and source-drain-gate electrodes. The E1−E4 electrodes were used for the characterization of electronic property for the $VO_2$ layer. Since there have been a few papers on the MI transition characteristics of $VO_2$ polycrystalline film, we have examined the material properties and examined the device characteristics using the VO$_2$ film on a glass substrate. The device structure with the 2.0-mm-square VO$_2$ channel was fabricated on an alkaline-free glass substrate (Corning® EAGLE XG®, substrate size: 10×10×0.7 mm$^3$) by pulsed laser deposition (PLD) using stencil masks.[17] A KrF excimer laser (wavelength of 248 nm, repetition rate of 10 Hz) was used to ablate ceramic target disks. The details of device fabrication are summarized in **Supplementary Material**. In order to fabricate the transparent device structure, which is essential to realize IR-transmittance modulation, all the films used in this study were selected from the wide-gap oxide materials. A nickel oxide/indium tin oxide (NiO/ITO) bilayer film was used as the transparent counter/top gate electrode, and F-doped SnO$_2$ film (SnO$_2$:F) was used as the source-drain and E1−E4 electrodes. The gate insulator consists of an amorphous 12CaO·7Al$_2$O$_3$ thin film with nanoporous structure (Calcium Aluminate with Nanopore, CAN);[15,18] since the 12CaO·7Al$_2$O$_3$ is a hygroscopic material, water vapor in air is automatically absorbed into the CAN film via the capillary action. Thus, a positive gate voltage ($V_g$) application between the gate and source electrodes induces electrochemical reactions such as protonation of the cathodic VO$_2$ layer (VO$_2$ + $x$H$^+$ + $x$e$^-$ → H$_x$VO$_2$) and hydroxylation of the anodic NiO layer (NiO + OH$^-$ → NiOOH + e$^-$)[19] (**Fig. 2(b)**). As a result, alternative positive and negative $V_g$ applications induce the reversible protonation / deprotonation of VO$_2$ channel, modulating it from IR-transparent insulator to IR-opaque metal.

**Figure 2(c)** shows the atomic force microscopy (AFM) image and reflection high energy electron diffraction (RHEED) pattern of the VO$_2$ film just deposited on a glass substrate. The film is composed of randomly oriented polycrystalline grains, which have the average grain size of 120 nm and the peak-to-valley value of 19 nm with the average surface roughness ($R_a$) of 2.3 nm. **Figure 2(d)** shows a bright-field scanning transmission electron microscopy (BF-STEM) image of the cross-section of the device prepared on the $VO_2$ film. The each grain size of $VO_2$ film was observed to be ~20 nm. Cosidering the average grain size in the AFM image, the $VO_2$ film is composed of many aggregation (with a diameter of 120 nm) of several 20-nm-size polycrystalline grains. The multi-layer structure of ITO (15 nm)/NiO (15 nm)/CAN (70 nm)/$VO_2$ (20 nm) was seen on a glass substrate. Crystal structure of the each layer was investigated by grazing incidence X-ray diffraction analyses (data not shown), which revealed that CAN and ITO layers were amorphous in nature expect for $VO_2$, NiO, $SnO_2$:F polycrystalline layers. Lighter spots with diameters of 10−20 nm are seen in the CAN film, indicating the presence of nanopores. The device structure was designed to keep a high transmittance in both IR and visible light region.

The conversion of electrical conductivity and IR transmittance was investigated at RT in ambient atmosphere (relative humidity value was ~30 % at 25 °C). The $V_g$ was applied between the gate and source electrodes of the device, where the drain electrode is open state and the gate current ($I_g$) flowing through NiO counter electrode /CAN gate insulator /$VO_2$ channel layer was in-situ measured during the $V_g$ application using a source measurement unit (Keithley 2450), as shown in Fig. 2(a). The opto-electronic properties were characterized immediately after each $V_g$ application, where the sheet resistance ($R_s$) was measured by the DC four probe method in the van der Pauw configuration and optical transmittance was measured by Ultraviolet-VIS/Near-IR microscope (Lamda 900s, PerkinElmer) and Fourier-transform IR spectrometer (FT-IR 660Plus, JASCO) with the light irradiation area of $0.2 \times 0.2$ cm$^2$.

We first evaluated the MI-phase modulation by applying $+V_g$ (protonation). **Figure 3** plots the $R_s$ of $VO_2$ channel as a function of applied $+V_g$, where the retention time at each $+V_g$ was set for 10 seconds. The $R_s$ was largely modulated from the virgin state (447 kΩ sq.$^{-1}$) by applying the $+V_g$; the $R_s$ exponentially decreased and their saturation was observed at $V_g \geq +8$ V, where the $R_s$ reached to 16 kΩ sq.$^{-1}$ at +12 V. The $R_s$ modulation should be mutually related with the flowing current in the device because of the electrochemical protonation. The inset of **Fig. 3** shows the relationship between the applied $+V_g$ and accumulated sheet electron density ($Q$), estimated as $Q = \frac{C}{A \times q}$, where $C$ is total coulomb amount calculated from the integral value of the $I_g$ versus retention time during the $+V_g$ application, $A$ is the surface area of $VO_2$ channel (0.2 × 0.2 cm$^2$), and $q$ is elementary charge, respectively. The $Q$ exponentially increased up to $1.0 \times 10^{17}$ cm$^{-2}$ ($\equiv 5.0 \times 10^{22}$ cm$^{-3}$) with respect to $+V_g$ and the slope becomes moderate at $V_g \geq +8$ V. The similar correlation between $Q$ and $R_s$ suggests that the $I_g$ flowing in the device originates from the ion current associated with electrochemical protonation of $VO_2$ channel and the critical $Q$ of MI switching is related with the ideal $Q$ value of $6.8 \times 10^{16}$ cm$^{-2}$ ($\equiv 3.4 \times 10^{22}$ cm$^{-3}$) required for the 100 % protonation of 20-nm-thick $VO_2$ layer, according to the following reaction: $VO_2$ + H$^+$ + e$^-$ ⇆ $HVO_2$ (the $Q$ is estimated by $Q = \frac{\rho \times t \times F}{M \times q}$, where $M$ is molar mass of $VO_2$, $\rho$ is the film density, $t$ is the film thickness, and $F$ is Faraday constant, respectively). This result suggests that almost all the provided electrons at $V_g$ up to +8 V were used for electrochemical protonation of the $VO_2$ channel, obeying Faraday's laws of electrolysis, and that the device operation can be controlled by the current density. It should be noted that the $Q$ continued to increase gradually at $V_g \geq +8$ V, while the $R_s$ was unchanged, suggesting that the $Q$ observed at $V_g \geq +8$ V originates from the gas formation by water electrolysis at the surfaces of cathodic VO$_2$ / anodic NiO layers. The protonated H$_x$VO$_2$ channel was stable under ambient conditions at RT after the $+V_g$ application; the $R_s$ of H$_x$VO$_2$ channel was unchanged for several days. Although it is necessary to test the retention-time dependence of the $R_s$ for H$_x$VO$_2$ channel kept under the several conditions, the result basically supports the non-volatility of device operation due to the electrochemical protonation.

**Figures 4(a) and 4(b)** sumarize the opto-electronic properties of the device. The temperature dependence of $R_s$ (**Fig. 4(a)**) was measured before and after applying $V_g$ of +12 V (protonation) and –30 V (deprotonation) for 10 seconds alternately at RT in air. The $R_s$ variation with respect to negative $V_g$ application is shown in **Supplementary Fig. S1**. The $R_s$–$T$ curves were measured up to 90 °C during the heating runs because the deprotonation of H$_x$VO$_2$ was previously confirmed to occur at $T$ = 100–150 °C.[16] The inset plots the temperature derivative curves of d(log $R_s$) / d$T$ to clearly visualize the $T_{MI}$. At the initial state, the MI transition was observed at $T_{MI}$ = 70 °C, which is defined as the peak position in d(log $R_s$) / d$T$ versus $T$, while it disappeared by applying $V_g$ = +12 V, indicating that the VO$_2$ channel changes from an insulator to a metal at RT, because of the decrease of $T_{MI}$ below RT by the protonation. The $R_s$–$T$ were reversibly modulated and recovered to initial state by applying $V_g$ = –30 V, where two orders of magnitude modulation of $R_s$ was observed at RT by the MI-phase conversion of VO$_2$.

It has been reported that the protonation of VO$_2$ (H$_x$VO$_2$) is thermodynamically favorable,[20] where hydrogen in VO$_2$ tends to form an strong O-H bond with the closest oxygen[12,21] and electron transfer from hydrogen onto the oxygen atom effectively reduce the electronegativity of the phase and makes it thermodynamically stable than that of pure $VO_2$ phase. Actually, deprotonation needed thermal annealing at higher temperature than that of protonation.[7] Therefore, the difference between the $+V_g$ (protonation) and $-V_g$ (deprotonation) should originate from the negative free Gibbs energy and activation barrier for the surface reaction, *i.e.* the in-diffusion and out-diffusion of $H^+$ transport have different interfacial resistances. Compared to the device of $VO_2$ epitaxial film on sapphire substrate,[16] the present device is operable by smaller DC voltage and shorter $V_g$ application time, suggesting that the polycrystalline surface of $VO_2$ film, shown in Fig. 2(c), enlarges the surface area with respect to that of the $VO_2$ epitaxial film and enables the effective protonation of $VO_2$ channel layer.

Then the optical transmission spectra (**Fig. 4(b)**) were measured. The initial device is transparent except for the weak absorption due to the transitions between the V 3d bands with crystal-field splitting at the wavelength ($\lambda$) > 500 nm.[22,23] and also due to the thin-film interference. By applying +12 V, it shows an abrupt transmittance decrease in the IR region, where the transmittance modulation ratio ($\Delta T$) at $\lambda$ = 3000 nm was 49 %, while almost no change is seen in the VIS region. These results indicate that the modulation from IR transparent insulator to IR opaque metal was successfully demonstrated by RT protonation.

We then measured the thermopower ($S$) of $VO_2$ channel protonated and deprotonated at each $\pm V_g$ in order to characterize the electronic-structure change resulting from carrier doping (protonation).[24] Since the $S$ basically reflects the energy differential of the density of states (DOS) around the Fermi level ($E_F$), $\left[\frac{\partial \text{DOS}(E)}{\partial E}\right]_{E=E_F}$, its value changes significantly due to the electronic-structure reconstruction across the $T_{MI}$. The $S$ was measured at RT by giving a temperature difference ($\Delta T$) up to ~4 K using two

Peltier devices, where the actual temperatures of both sides of VO$_2$ channel layer were monitored by two tiny thermocouples with tip diameter of 150 μm. The schematic measurement setup for thermopower was reported in Ref. 18. The thermo-electromotive force (Δ$V$) and Δ$T$ were simultaneously measured, and the $S$ were obtained from the linear slope of the Δ$V$–Δ$T$ plots. **Figure 5(a)** shows the relationship between $S$ and 1/$R_s$, where the positive $V_g$ up to +12 V in a +1 V step was first applied for protonation and then negative $V_g$ up to –30 V in a –3V step was applied for deprotonation. The $S$ were always negative, indicating that the H$_x$VO$_2$ layer is an n-type conductor. The |$S$| linearly decreased from 420 μV K$^{-1}$ to 30 μV K$^{-1}$ with logarithmic increase in 1/$R_s$ was reversibly observed for the application of ±$V_g$, suggesting that protonation of the VO$_2$ channel provides electrons to the conduction band, and the energy derivative of DOS near the $E_F$ becomes moderate, resulting in the consequent reduction of |$S$|.

Here we like to compare the present results with another electron-doping system, (V$_{1-y}$W$_y$)O$_2$ polycrystalline films grown on glass substrates. **Supplementary Figs. S2 and S3** summarize the opto-electronic properties of (V$_{1-y}$W$_y$)O$_2$ films, where the MI transition was also modulated by W doping and the $T_{MI}$ was suppressed below RT at $y$ = 0.06 (**Fig. S2**). The |$S$| of (V$_{1-y}$W$_y$)O$_2$ film decreased with increasing $y$ and became constant at small |$S$| of 30 μV K$^{-1}$ (**Fig. S3**), which is the same with the present metallic H$_x$VO$_2$ film (30 μV K$^{-1}$).

Lastly, in order to analyse the device operation, a simple bi-layer model of thermopower was applied to estimate the thickness ($d$) of metallic H$_x$VO$_2$ layer (**Fig. 5(b)**), where the parallel circuit composed of metal (M) and insulator (I) layers was considered to calculate the combined thermopower. Since the thermopower depends on both the conductivity and thickness of each layer, the observable |$S$| can be expressed by

the equation of $|S|_{obs.} = (\sigma_{sM} \cdot |S|_M + \sigma_{sI} \cdot |S|_I)/(\sigma_{sM} + \sigma_{sI})$, where sheet conductance $\sigma_{sM}$ and $\sigma_{sI}$ are defined as $\sigma_M \times d$ and $\sigma_I \times (20-d)$, respectively. The physical properties of $\sigma$ and $|S|$ for the I and M phases are used from those of virgin VO$_2$ film and $(V_{1-y}W_y)O_2$ ($y = 0.06$) film, respectively. The $d$ gradually increases from the surface[25] and finally all of the channel region changes to metallic state by applying $+V_g$. The $d$ reversely modulated by applying $-V_g$, indicating that the metallized H$_x$VO$_2$ film region can be controled by applied $V_g$.

In summary, we have demonstrated the IR-transmittance tunable MI conversion device, which has a three-terminal TFT geometry consisting of transparent oxide thin films of VO$_2$ active channel, water-leakage-free CAN gate insulator, NiO counter layer / ITO gate electrode, and SnO$_2$:F source-drain electrodes, on a glass substrate. At initial state, the device was insulator and transparent in the IR region. For $+V_g$ application, the $R_s$ decreased due to the protonation of VO$_2$ channel and the device became IR opaque state. For $-V_g$ application, deprotonation of H$_x$VO$_2$ channel occurred and the device returned to insulator / IR transparent state. The two orders of magnitude modulation of $R_s$ and 49 % modulation of IR transmittance at $\lambda$ of 3000 nm was simultaneously demonstrated at RT by the metal-insulator phase conversion of VO$_2$ in a non-volatile manner.

The present IR-transmittance tunable MI conversion device has several advantages. The device can be fabricated on a glass substrate, which is suitable for the application to glass window; the device fully transmits IR in the OFF state, whereas it does not transmit in the ON state. Meanwhile, the device can function as ON/OFF power switch for electronic device to control the in-house temperature. Moreover, the device can be operated by RT-protonation without sealing thanks to the water-leakage-free CAN gate insulator; the all-solid-state structure can resolve the liquid-leakage problem, which is a beneficial point compared to the liquid-electrolyte gated devices.[26] Although the demonstration of power saving by this device should be necessary to show the suitability for the practical application, the present device concept provides a potential gateway to a new functional device for future energy saving technologies such as advanced smart windows.

**Supplementary Material**

See http://dx.doi.org/XXXX for Supplementary information on device fabrication and sample preparation for transmission electron microscopy, and Supplementary discussion on device operation mechanism and the protonation of $VO_2$ channel layer, and Supplementary Figs. S1–S3.

The authors thank N. Hirai for experimental help on TEM/STEM analyses. The TEM/STEM analyses, conducted at Hokkaido University, were supported by Nanotechnology Platform Program from MEXT. This work was supported by Grant-in-Aid for Scientific Research A (25246023), Grant-in-Aid for Young Scientists A (15H05543), Grant-in-Aid for Challenging Exploratory Research (16K14377), and Grant-in-Aid for Scientific Research on Innovative Areas (25106007) from JSPS. T.K. was supported by PRESTO, JST (JPMJPR16R1).

**Figure captions**

**FIG. 1.** Concept of an IR-transmittance tunable MI conversion device. The device can control IR transmittance, while maintaining VIS transparency, and electrical conductivity at the same time. In the ON state (right figure), the IR cannot be transmitted through the device, whereas it can be transmitted in the OFF state (left figure), which can work as the thermal radiation control from sunlight. Further, the device can function as an ON/OFF power switch for electronic devices to control the in-house temperature. Such device on a glass substrate would be useful as an energy-saving smart window application.

**FIG. 2.** (a) Schematic device structure with three-terminal transistor geometry consisting of $VO_2$ active layer, water-leakage-free CAN gate insulator, NiO counter layer / ITO gate electrode, and $SnO_2$:F source-drain electrodes, on a glass substrate. E1−E4 electrodes of $SnO_2$:F film were used for the electrical transport measurements. (b) Device operation mechanism. During the positive $V_g$ application, protonation of $VO_2$ layer and hydroxylation of NiO layer occur simultaneously. Conversely, $H_xVO_2$ and NiOOH return to $VO_2$ and NiO during the negative $V_g$ application. (c) Topographic AFM image of $VO_2$ film. RHEED pattern is superimposed in the figure, confirming the polycrystalline nature of $VO_2$ film. (d) Cross-sectional BF-STEM image of the device. Trilayer structure composed of $VO_2$ (20 nm), CAN (70 nm), and NiO (15 nm) / ITO (15 nm) layers is seen. Lighter spots in the CAN layer indicate nanopores, which is fully occupied with water.

**FIG. 3.** Sheet resistance ($R_s$) as a function of $V_g$ up to +12 V, where $R_s$ was measured after holding the $V_g$ application for 10 seconds at each step. Inset shows $V_g$ dependence of sheet electron density ($Q$) accumulated during $V_g$ application. The $R_s$ decreased with increasing $V_g$ up to +8 V, where $Q$ also exponentially increased with $V_g$ due to the electrochemical protonation of $VO_2$ channel.

**FIG. 4.** (a) Temperature dependence of $R_s$ measured before (red closed circles) and after applying $V_g$ alternately at +12 V (protonation, blue closed triangles) and −30 V

(deprotonation, red open diamond symbols). Each curve was measured after the $V_g$ turned off, where the $V_g$ application time was 10 seconds. Inset plots the temperature derivative curves of $d(\log R_s) / dT$. (b) Optical transmittance spectra measured before (red line) and after (blue line) applying $V_g = +12$ V. The optical transmittance modulation ratio at $\lambda$ of 3000 nm was 49 %.

**FIG. 5.** Thermopower analysis of the device operation. (a) Thermopower ($S$) as a function of $1/R_s$ at RT by applying $+V_g$ (protonation, red circles) and $-V_g$ (deprotonation, blue triangles). The linear relation between $S$ and logarithmic $1/R_s$ was reversibly observed by $\pm V_g$ application. (b) The thickness ($d$) of metallic $H_xVO_2$ layer with respect to $1/R_s$, estimated by thermopower analysis. The $d$ increases from the surface and finally all of the channel region changes to metallic state by applying $+V_g$. The $d$ reversely modulated by applying $-V_g$.

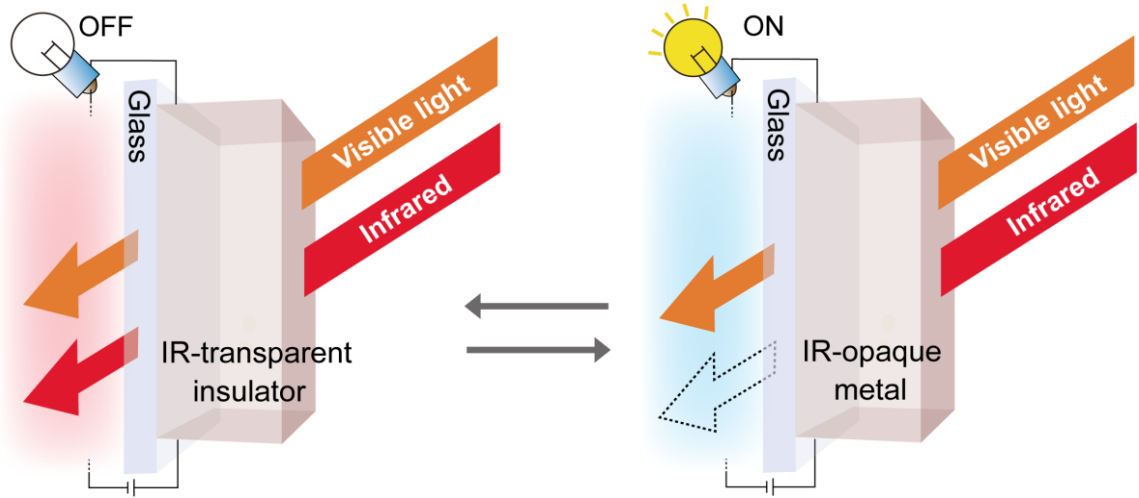

Figure 1

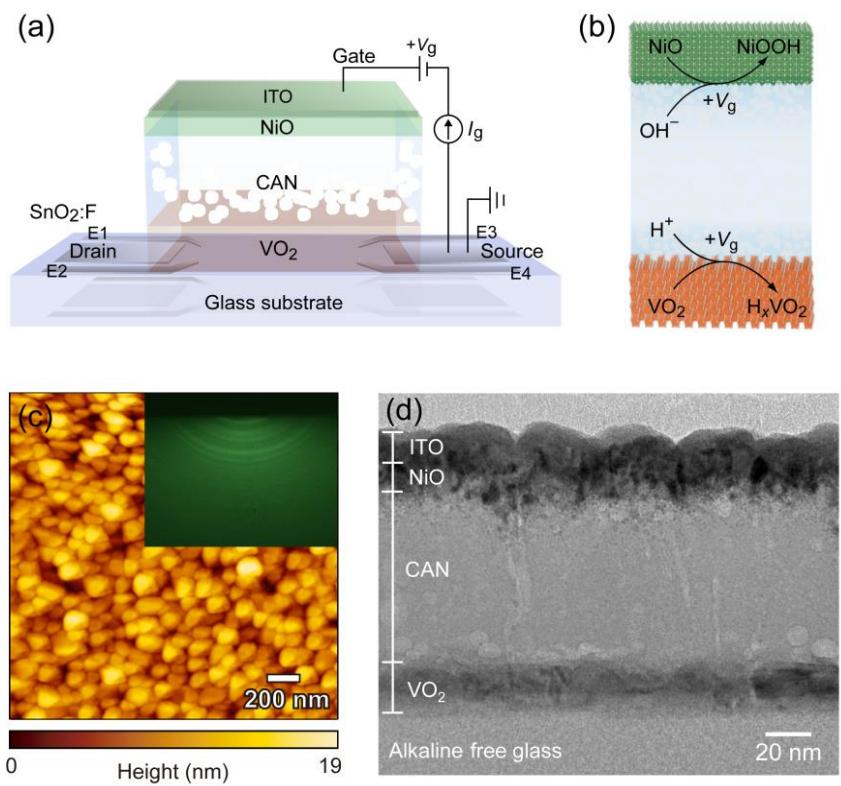

Figure 2

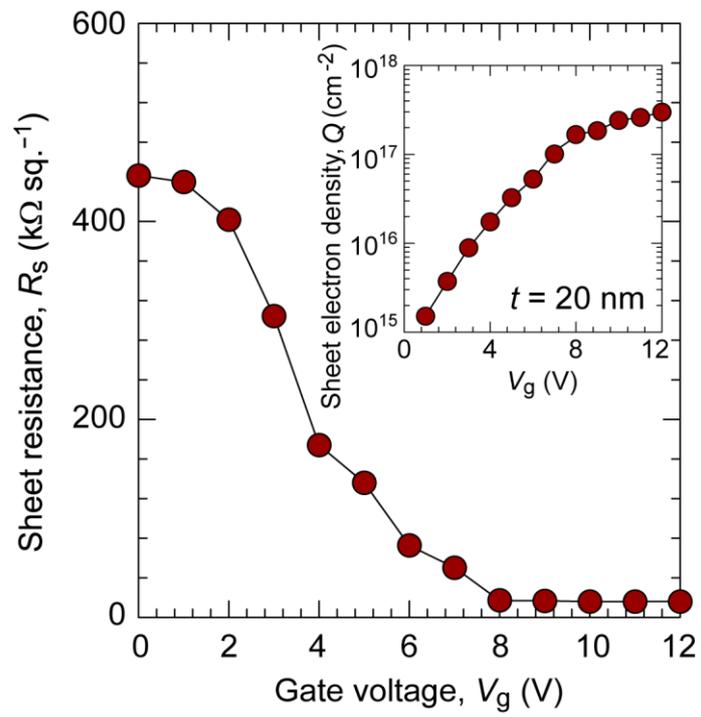

Figure 3

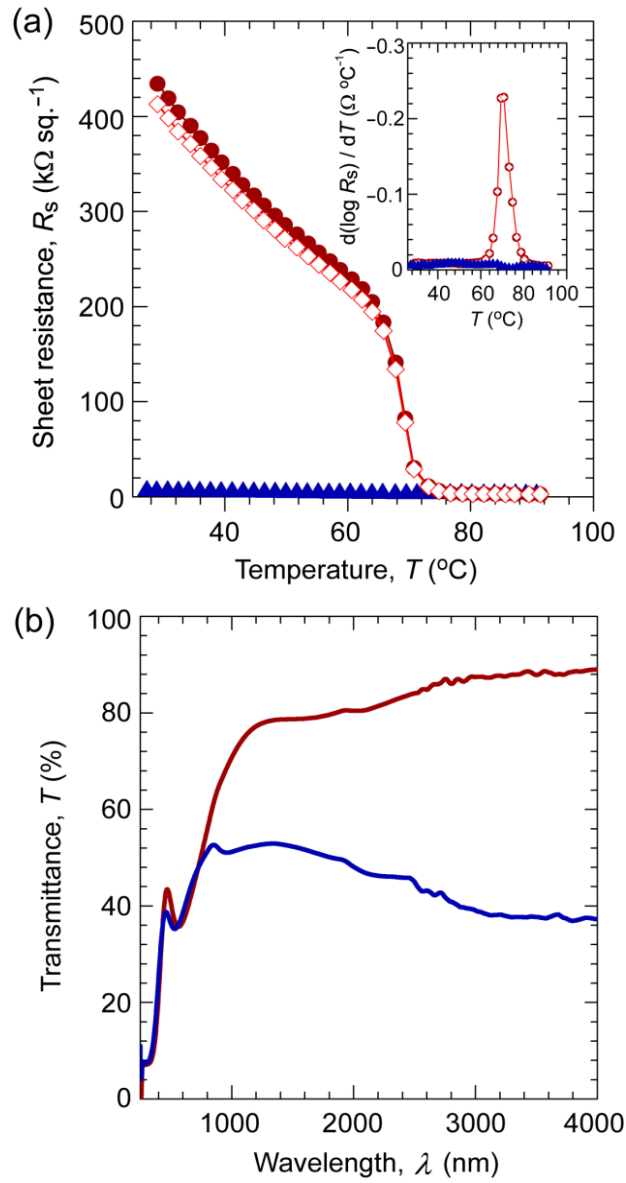

Figure 4

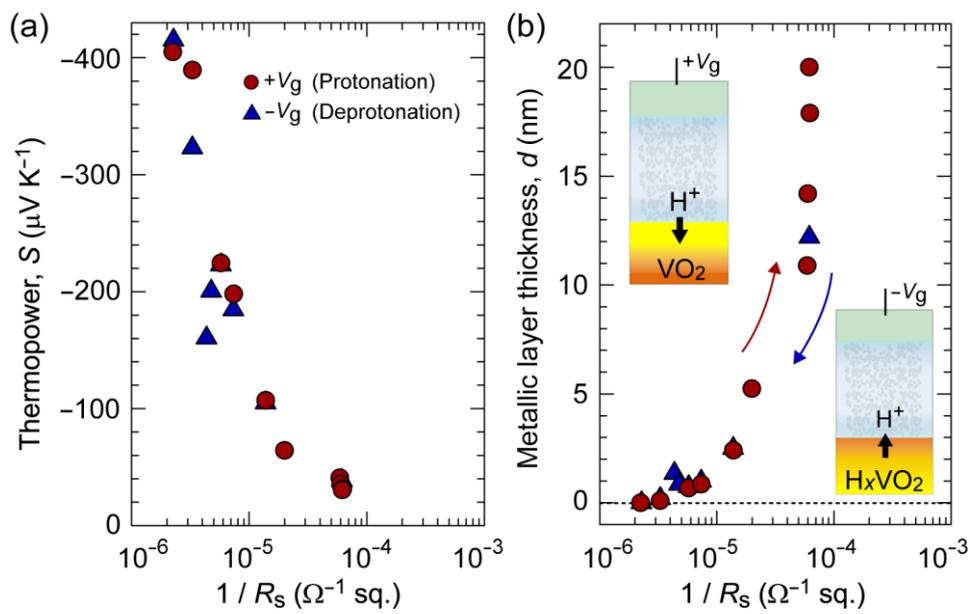

Figure 5